\documentclass[final,5p,twocolumn,sort&compress]{elsarticle}

\usepackage{amsmath}
\usepackage{amssymb}
\usepackage{units}
\usepackage{slashed}

\newcommand{\Tr}{\mathrm{Tr}}

\journal{Physics Letters B}
\begin{document}

\begin{frontmatter}

\title{Probing the quark-gluon interaction with hadrons}

\author{H\`elios Sanchis-Alepuz}
\ead{helios.sanchis-alepuz@physik.uni-giessen.de}
\author{Richard Williams}
\ead{richard.williams@physik.uni-giessen.de}

\address{Institut f\"ur Theoretische Physik, Justus-Liebig--Universit\"at Giessen, 35392 Giessen, Germany.}

\begin{abstract}
We present a unified picture of mesons and baryons in the 
Dyson-Schwinger/Bethe-Salpeter approach, wherein
the quark-gluon and quark-(anti)quark interaction follow from a systematic 
truncation of the QCD effective action and includes all its tensor structures.
The masses of some of the ground state mesons and baryons are found to be in 
reasonable 
agreement with the expectations of a `quark-core calculation', suggesting a
partial insensitivity to the details of the quark-gluon interaction. However, 
discrepancies remain 
in the meson sector, and for excited baryons, that suggest higher order corrections 
are relevant and should be investigated following the methods outlined herein. 
\end{abstract}


\end{frontmatter}


\section{Introduction}

%
%
%
%

Hadrons provide a rich experimental environment for the study of the 
strong interaction, from details of the resonance spectrum to 
form factors and transition decays via electromagnetic probes. These reflect 
the underlying substructure of bound states by resolving, in a
non-trivial way, the quarks and gluons of which they are composed. A 
theoretical understanding of hadrons in terms of these underlying degrees of 
freedom, interacting as dictated by quantum chromodynamics (QCD), is an 
on-going effort. Many approaches tackle it in different ways, simplifying 
certain aspects of the theory. Probing sensibly our theoretical constructs with 
experimental input thus provides understanding of the theory itself.

In continuum approaches to QCD, it is not possible in 
general to include all possible correlation functions in a calculation, as 
there 
are infinitely many of them.
Although this can be viewed as a limitation of continuum 
approaches, only a finite number of these correlation 
functions have a significant role in the observable properties of hadrons.
Therefore, by including a greater number of relevant correlation functions into the system, 
continuum methods provide an ideal framework to unravel the underlying mechanisms that 
generate observable effects from the elementary and non-observable degrees of freedom of QCD. 
This is in contrast to lattice QCD
calculations, which can be viewed as theoretical experiments in the sense
that, although they contain \emph{a priori}  all the dynamics of QCD, it is 
challenging to single out individual contributions to a particular 
measurement. This makes these two approaches complementary.

Amongst the different continuum approaches, the combination of Dyson--Schwinger 
(DSE) and Bethe--Salpeter equations (BSE) has proven to be extremely useful in 
the calculation of hadronic properties from 
QCD~\cite{Bashir:2012fs,Eichmann:2013afa,Cloet:2013jya}. Typically, solutions 
of 
DSEs constitute the building blocks (propagators and vertices) of bound-state 
calculations using BSEs, which provide the bridge between QCD and observables.
As described in more detail below, the interaction terms that are kept in the 
DSE determine the interaction kernels among constituents in the 
BSEs, thereby defining a particular truncation of the DSE/BSE system. One works 
towards a model independent truncation by including a larger set of interaction 
terms; although this programme is obviously not achievable in its 
totality, it is expected that there will be some degree of convergence at the level of this vertex expansion. 

Here we focus upon the inclusion of the quark-gluon interaction, the reliable construction of which is 
a challenging task. However, one is guided by various symmetries -- notably that of 
chiral symmetry -- that provide 
for stringent constraints. To implement these symmetries at the level of the quark and 
gluon interaction, simplifications are clearly necessary which typically fall into 
three categories: (i)~the quark-gluon vertex is 
truncated to its tree-level component times a momentum-dependent effective coupling with 
the quark DSE and hadron BSEs solved self-consistently 
\cite{Maris:1997tm,Maris:1999nt};
(ii)~a more sophisticated model for the quark-gluon vertex is used, with the 
contribution from its different tensor structures modelled, hence abandoning 
self-consistency but gaining instead flexible insight into the relative 
importance of each of these structures~\cite{Chang:2009zb,Chang:2010hb,Chang:2011ei,Heupel:2014ina}; 
(iii)~some non-perturbative effects of the quark and gluon interaction are 
taken into 
account by solving the quark-gluon vertex DSE self-consistently, but 
potentially 
introducing some truncation artifacts~\cite{Watson:2004kd,Fischer:2009jm,Williams:2014iea}. 
We follow here the latter approach, since as we demonstrate in this letter it
enables the controlled inclusion of interaction mechanisms based on a loop expansion 
of the effective action.

The effective action is a generating functional for proper vertex functions, and may
be considered as a means to define the quantum field theory given an action, since all necessary Green's 
functions can be derived from it. Related are the n-particle irreducible (nPI) effective actions that form a family 
of different, but equivalent, representations of the same generating 
functional (see, e.g.~\cite{Berges:2004pu}). They are defined 
as functionals of all $m\le n$ Green's functions of the theory (fields, 
propagators, vertices, etc.). Although its exact form is not known in general, 
its loop expansion in $\hbar$ is well-defined~\cite{Berges:2004pu} and can in 
practice be performed. Moreover, each term in the expansion already captures both 
perturbative and non-perturbative physics.

One reason that makes nPI techniques a powerful tool is that they provide a natural link 
between bound-state equations described by BSEs and the 
propagators and vertices provided by DSEs~\cite{Fukuda:1987su,Komachiya:1989kc,McKay:1989rk,Carrington:2010qq,Sanchis-Alepuz:2015tha}.
A truncation of the loop expansion at a certain order translates into a unique 
prescription for the truncation of the DSEs and the BSEs that maintains symmetries. For the study 
of two- and three-body states, it suffices to use either the 2PI effective 
action, which is defined in terms of fully-dressed propagators but bare 
vertices~\cite{Cornwall:1974vz}, or the 3PI effective action, which is defined 
also in terms of the fully-dressed vertices.
In this work we restrict ourselves to the 2PI case and defer the use of the 3PI 
effective action to a future and more comprehensive study.

In connection with the three categories outlined above, it is 
worth mentioning here that the somewhat hybrid possibility of supplementing some of QCDs 
degrees of freedom in favor of effective ones, such as pions, has also been 
explored~\cite{Hecht:2002ej,Fischer:2007ze,Fischer:2008sp,Fischer:2008wy,
Sanchis-Alepuz:2014wea,Braun:2014ata}. These can be viewed as approximate representations of 
the four-quark vertex in the 4PI formalism that introduces at the first step
decay channels and a mixing with tetraquark states.

In the present work, we incorporate the results of a recent study of the 
quark-gluon vertex from a truncated DSE~\cite{Williams:2014iea} in the 
calculation of 
meson and baryon masses. That truncation can be interpreted in the context of the 2PI effective action at 3-loop. 
Although on the technical side this is no novelty for 
meson
calculations \cite{Fischer:2009jm,Williams:2009wx}, it is the first time that corrections 
incorporating the gluon self-interaction have have been included in the  
covariant three-body baryon calculation.
While there exist other recent investigations of the quark-gluon 
vertex~\cite{Hopfer:2013np,Aguilar:2014lha,Mitter:2014wpa}, 
these have not yet been confronted with the challenge of reproducing hadron 
phenomena for reasons we discuss below.

Finally, we wish to stress that this work represents only the first step in an 
on-going effort to incorporate realistic QCD's Green's functions into the 
self-consistent study of bound states. Not surprisingly, the low-order of the 
truncation used performs only as well as simple phenomenological models such as rainbow-ladder. However, but
most importantly, it serves as a proof of principle that such an endeavor is feasible, as further increasing 
the order of the truncation does not increase the technical complexity dramatically.

\section{Framework}

%
%
%
%

%
\begin{figure}[!t]
\begin{center}
\includegraphics[height=1.75cm]{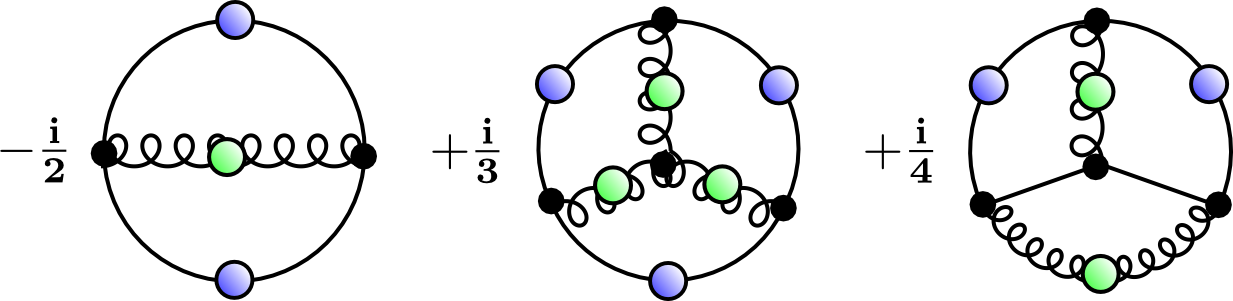}
\caption{The 2-particle irreducible term $\Gamma_2[\Psi,G]$ in the definition of 
the effective action, up to three loops.}
\label{fig:eff_action}
\end{center}
\end{figure}
\begin{figure}[!t]
\begin{center}
\includegraphics[height=1.0cm]{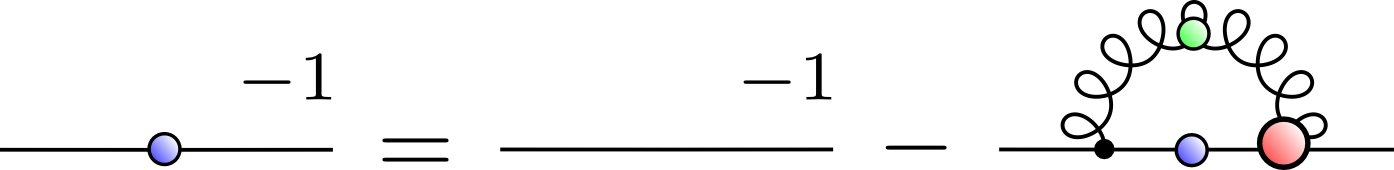}
\caption{The Dyson--Schwinger equation for the quark propagator.}
\label{fig:quarkdse}
\end{center}
\end{figure}

The starting point for the study of hadronic observables in the present 
framework is thus the effective action
\begin{align}\label{eq:eff_action}
        \Gamma[\Psi,G]=\textrm{S}[\Psi]+i\Tr\log G-i\Tr 
G_0^{-1}G+\Gamma_2[\Psi,G]~,
\end{align}
where $S$ is the classical action, and $\Psi$ and $G$ collectively represent the fields and full 
propagators of QCD, respectively. The term $\Gamma_2[\Psi,G]$ contains 
two-particle irreducible diagrams only and $G_0$ denotes the classical 
propagators. To proceed, we perform a loop expansion of $\Gamma_2[\Psi,G]$ to 
three-loop order, as shown in Fig.~\ref{fig:eff_action}. Moreover, we keep only 
the non-Abelian term that connects the gauge to the matter sector and neglect 
the Abelian correction (third diagram), as it is expected to be subleading in the large-$N_c$ 
limit; whether this is indeed the case for the description of hadron phenomena 
must certainly be tested and will be the subject of future work.

\begin{figure}[!t]
\begin{center}
\includegraphics[scale=0.5]{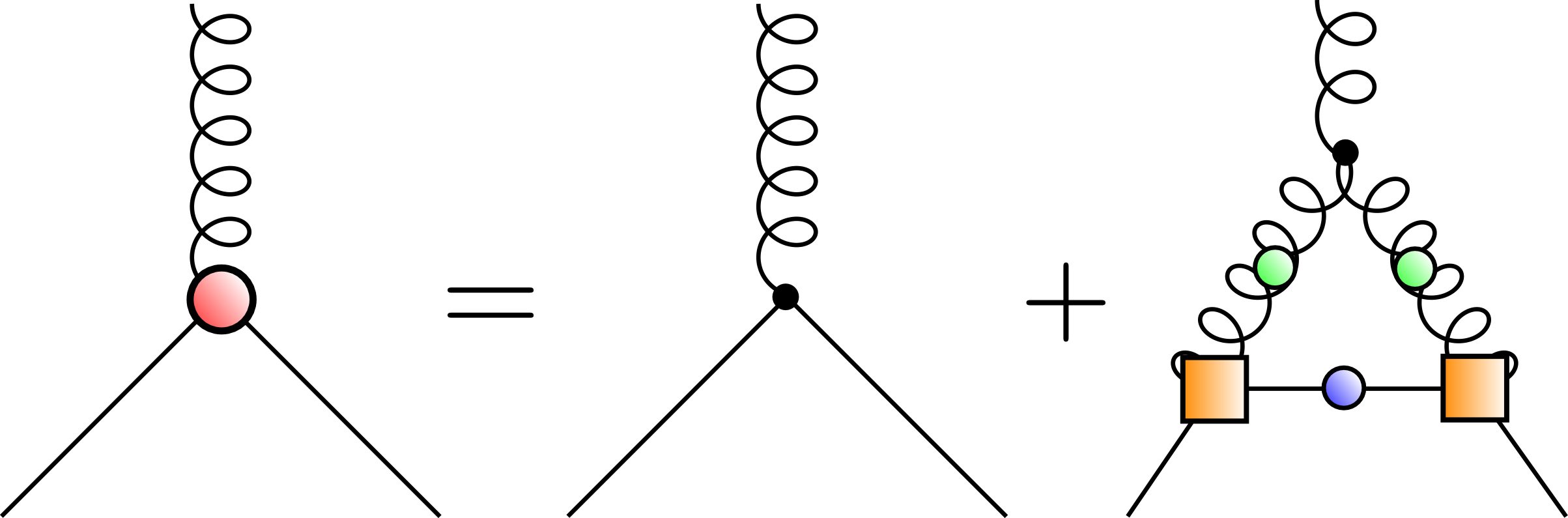}
\caption{The truncated DSE for the quark-gluon vertex, showing the non-Abelian 
correction.
Boxes represent an RG improvement of the bare vertex; 
see \eqref{eq:vertex_NA_corrections}.}\label{fig:qgvertextrunc}
\end{center}
\end{figure}

The next basic element in meson and baryon calculations is the fully-dressed 
quark propagator. It is given by the quark DSE, see Fig.~\ref{fig:quarkdse}
\begin{align}\label{eq:quarkdse}
	S^{-1}(p;\mu) & = Z_2S_0^{-1}(p) + \Sigma(p;\mu) \;,
\end{align}
with quark self-energy
\begin{align}\label{eq:selfenergy}
	\Sigma(p;\mu) = g^2 Z_{1f}C_F\int_k \gamma^\mu S(q)\Gamma^\nu(q,p)D_{\mu\nu}(k)\;.
\end{align}
Here $q=k+p$, the integral measure is $\int_k=\int d^4k/(2\pi)^4$ and $Z_{1f}$, 
$Z_2$ are renormalization constants for the quark-gluon vertex and quark propagator respectively. It is clearly dependent upon both the 
gluon 
propagator $D_{\mu\nu}(k)$ and the quark-gluon vertex $\Gamma^\nu(q,p)$. The 
(Landau gauge) propagators are
\begin{align}
	S^{-1}(p) &= Z_f^{-1}(p^2)\left( i\slashed{p} + M(p^2)\right)\;, \\
	D^{\mu\nu}(k) &= T^{\mu\nu}_{(k)} Z(k^2)/k^2\;,
\end{align}
with quark wavefunction $Z_f^{-1}(p^2)$, dynamical mass $M(p^2)$ and gluon 
dressing $Z(k^2)$. The transverse projector is 
$T^{\mu\nu}_{(k)} = \delta^{\mu\nu} - k^\mu k^\nu / k^2$. From the 2PI 
effective action, the quark DSE is determined via a functional derivative with 
respect to the quark propagator
\begin{align}
    \Sigma=-i\frac{\delta \Gamma_2}{ \delta G}\,.
\end{align}
The expansion of the effective action in Fig.~\ref{fig:eff_action} thus defines 
a truncation of the quark DSE, which is equivalent to the truncation of the 
quark-gluon vertex DSE shown in Fig.~\ref{fig:qgvertextrunc}. Specifically, the 
truncated vertex DSE can be given as a summation of vertex corrections 
\begin{align}\label{eq:vertex_corrections}
	\Gamma^\mu(l,k) = Z_{1f}\gamma^\mu + \Lambda^\mu_{\textrm{NA}} +\ldots\,,
\end{align}
with $\Lambda^\mu_{\textrm{NA}}$ the 
non-Abelian term (see Fig.~\ref{fig:qgvertextrunc}), and 
the ellipsis denotes contributions not considered here. 
The non-Abelian correction has the form
\begin{align}\label{eq:vertex_NA_corrections}
	\Lambda^\mu_{\textrm{NA}}(l,k)=\frac{N_c}{2} 
\int_q&\widetilde{\Gamma}_\alpha\left(l_1,-q_1\right) S(q_3) 
\widetilde{\Gamma}_\beta\left(l_2,-q_2\right) \\
    \times& \Gamma^{\alpha^\prime\beta^\prime\mu}_{3g}(q_1,q_2,p_3) 
D^{\alpha\alpha^\prime}(q_1)D^{\beta\beta^\prime}(q_2)\;,\nonumber
\end{align}
where the internal vertices $\widetilde{\Gamma}^\mu$ retain only the tree-level 
tensor structure and are supplemented by an enhancement factor in the infrared. 
Such an enhancement is needed in order to account for the effect 
on the quark-gluon vertex of higher loop terms in the 2PI effective action. An 
alternative to 
this is using a 3PI or higher effective action to define the truncation, which 
means that both vertices and propagators are fully dressed and independent 
objects~\cite{Komachiya:1989kc,Sanchis-Alepuz:2015tha}. Further details on 
\eqref{eq:vertex_NA_corrections} and its solution can be found 
in~\cite{Williams:2014iea}.

A second functional derivative of the effective action with respect to the 
quark propagator defines the quark-antiquark BSE interaction kernel 
$K^{q\bar{q}}$~\cite{Fukuda:1987su,Komachiya:1989kc,McKay:1989rk,Sanchis-Alepuz:2015tha,Cornwall:1974vz}. 
Equivalently,
\begin{align}\label{eq:cutting_lines}
	\big[K^{q\bar{q}}\big]_{ik;lj}=
    -\frac{\delta}{\delta \big[S(x,y)\big]_{kl}} 
\big[\Sigma\left(x^\prime,y^\prime\right)\big]_{ij}\;\;,
\end{align}
followed by a Fourier transform to momentum space. This is the kernel that 
appears in the BSE description of a meson as a 
quark-antiquark bound state
\begin{align}\label{eq:mesonbse}
	\big[\Gamma_M(p, P)\big]_{ij} = \int_k\big[K^{(q\bar{q})}\big]_{ik;lj} 
\big[\chi_M(k,P)\big]_{kl} \;,
\end{align}
where $\Gamma_M(p, P)$ is the Bethe--Salpeter amplitude and 
$\chi_M(k,P)=S(k_+)\Gamma_M(k,P)S(k_-)$ its wavefunction. The quantum numbers 
of 
the state are defined by its covariant 
decomposition~\cite{LlewellynSmith:1969az,Krassnigg:2010mh,Fischer:2014xha}. 
This gives access to both the mass of the bound-state as well as details of 
its internal structure. 

The crucial motivation for such a definition of the 
Bethe--Salpeter kernel is that chiral symmetry, as expressed via the 
axial-vector Ward--Takahashi identity (axWTI)
\begin{align}\label{eq:axWTI}
	\big[\Sigma(p_+)\gamma_5 & + \gamma_5\Sigma(p_-)\big]_{ij} = \\
                    &\int_k \big[K^{(q\bar{q})}\big]_{ik;lj} 
\big[S(k_+)\gamma_5  + \gamma_5 S(k_-)\big]_{kl}\;,\nonumber
\end{align}
is correctly implemented in the calculation of meson 
properties~\cite{Munczek:1994zz}. This guarantees, in particular, the 
identification of the pion as a Goldstone boson in the chiral limit.
\begin{figure}[!t]
\begin{center}
\includegraphics[scale=0.5]{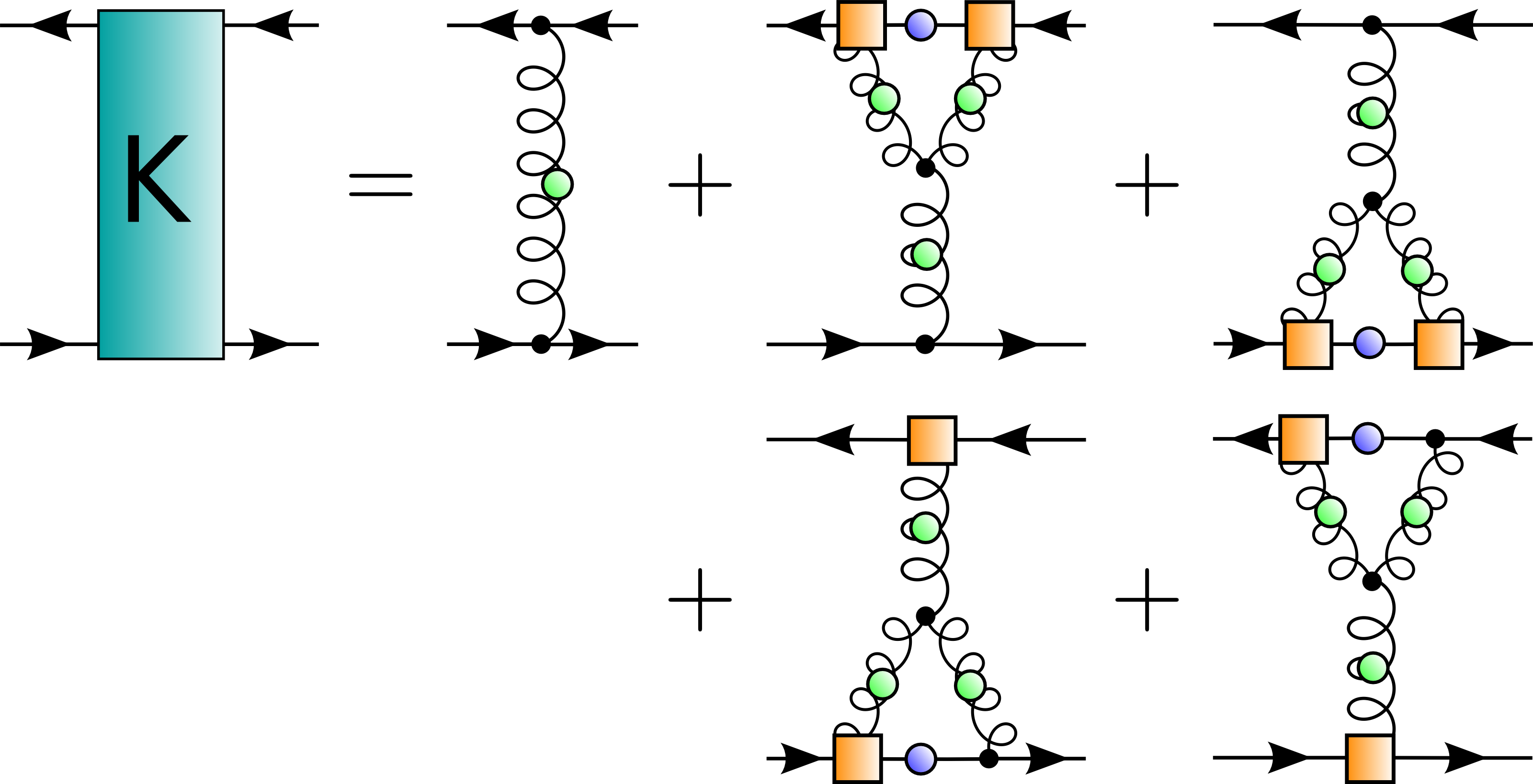}
\caption{The quark-antiquark kernel beyond 
rainbow-ladder.}\label{fig:kernel2body}
\end{center}
\end{figure}

With the truncation of the vertex DSE given in 
\eqref{eq:vertex_NA_corrections}, the cutting process gives rise to a 
quark-(anti)quark kernel, given in Fig.~\ref{fig:kernel2body}, whose
topology is analogous to a gluon ladder
\begin{align}\label{eqn:kernelqqbarbrl}
	\big[K^{(q\bar{q})}\big]_{ij;mn} &= 
D_{\mu\nu}(q)\bigg[\big[\gamma^\mu\big]_{ij}  \big[\gamma^\nu\big]_{mn} \\
                                 &+ \big[\gamma^\mu\big]_{ij}  
\big[\Lambda^\nu\big]_{mn} + \big[\Lambda^\mu\big]_{ij} 
\big[\gamma^\nu\big]_{mn}\bigg]\;. \nonumber
\end{align}
This, taken with the truncated quark-gluon vertex above, satisfies chiral 
symmetry by construction. Further corrections to the 
quark-gluon vertex, in 
particular those featuring additional quark propagators, yield different 
topologies and/or are higher loop order; they are beyond the scope of the 
present work but can in principle be included.

It is important to stress that when the 2PI effective action is used to define 
the truncation at the level of vertex functions, the expansion must be chosen 
such that the functional derivative \eqref{eq:cutting_lines} can be formally 
performed to define the chiral-symmetry preserving BSE kernels. This implies 
that it must be possible to resolve diagrammatically the quark lines in the 
vertex corrections~\eqref{eq:vertex_corrections}. This is the reason why 
quark-gluon vertices defined as ans\"atze are difficult to accommodate (and, 
hence, test) consistently in this framework.

It is illustrative to consider here the case of the two-loop expansion of 
the 2PI effective action, which leads to the well-known rainbow-ladder truncation. There, 
suppressing constants and color factors for simplicity, we have
\begin{align}  
	\big[\Sigma(p)\big]_{ij} \simeq \int_k \big[\gamma^\mu 
S(q)\gamma^\nu\big]_{ij} D_{\mu\nu}(k)\;,
\end{align}
from which a functional derivative provides the kernel
\begin{align}\label{eqn:ladderkernel}
	\big[K^{(q\bar{q})}\big]_{ik;lj} \simeq \big[\gamma^\mu\big]_{ik} 
\big[\gamma^\nu\big]_{lj} D_{\mu\nu}(q)\;.
\end{align}
In practical calculations the gluon propagator 
is modelled by a phenomenological function which includes the aforementioned 
infrared enhancement of the quark-gluon vertex from (omitted) higher order corrections.

Since we wish to also consider bound-states of three-quarks, it proves useful 
as 
an intermediate step to formulate the diquark bound-state
\begin{align}\label{eqn:diquarkbse}
	\big[\Gamma_D(p, P)\big]_{ij} = \int_k\big[K^{(qq)}\big]_{ik;lj} 
\big[\chi_D(k,P)\big]_{kl} \;,
\end{align}
with diquark amplitude $\Gamma_D(p, P)$ and wavefunction 
$\chi_D(k,P)=S(k_+)\Gamma_D(k,P)S^T(-k_-)$. The superscript $T$ denotes 
transposition; the covariant decomposition of a $J^{-P}$ diquark $\Gamma_D$ are
that of a $J^P$ meson $\Gamma_M C$ together with a charge conjugation matrix. 
The corresponding diquark kernel is then 
\begin{align}\label{eqn:diquarkkernel}
	\big[K^{(qq)}\big]_{ik;lj} =\big[K^{(q\bar{q})}\big]_{ik;jl} \;,
\end{align}
that is, it amounts to a transposition of the lower spin-line.
For general ladder-like kernels, such as that in \eqref{eqn:ladderkernel} the 
color factors are $(N_c^2-1)/2N_c$ and $-(N_c+1)/2N_c$ for a meson and diquark, 
respectively.

\begin{figure}[!t]
\begin{center}
\includegraphics[scale=0.45]{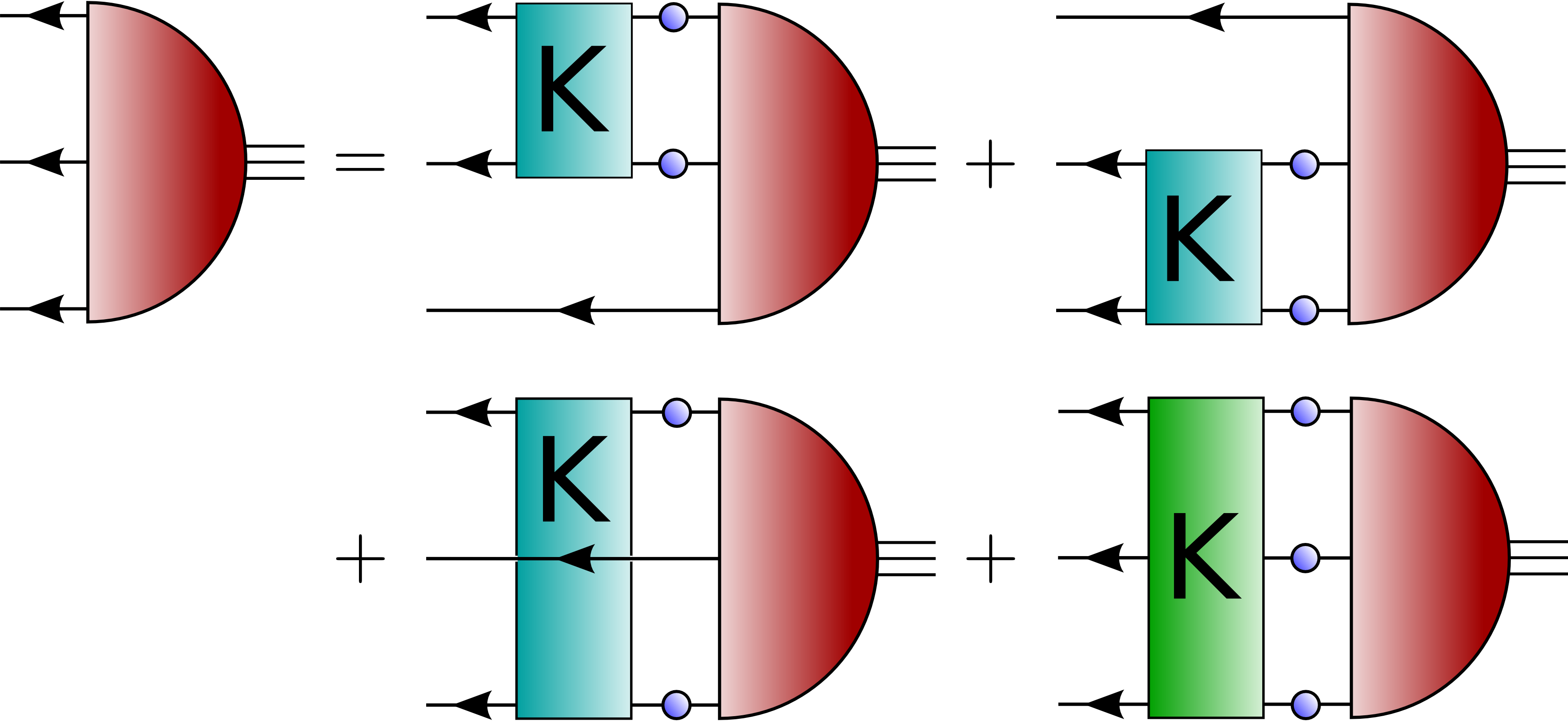}
\caption{The covariant Faddeev equation for the baryon, see 
~\eqref{eq:3bBSEcompact}.}\label{fig:baryonbse}
\end{center}
\end{figure}

%
%
The analogue of the Bethe--Salpeter equation for baryons can now be formulated. 
It contains the permuted sum of the two-body quark-quark kernel $K^{(qq)}$ and 
an irreducible three-body kernel $K^{(qqq)}$ ~\cite{Komachiya:1989kc,Eichmann:2009qa}
\begin{align}\label{eq:3bBSEcompact}
	\Psi = \big[K^{(qqq)}\big] G_0^{(3)}\Psi + \sum_{a=1}^3 
\big[K^{(qq)}_{a}\big]~S^{-1}_{a}~G_0^{(3)}\Psi\,,
\end{align}
see Fig.~\ref{fig:baryonbse}. We use here a compact notation, omitting indices, 
where 
implied discrete and continuous variables are summed or integrated over, 
respectively. Here $G_0^{(3)}$ represents the product of three fully-dressed 
quark propagators $S$. The subscript $a$ labels the quark spectator to the
two-body interaction. At the truncation level of the 2PI effective action we 
will consider in this 
work, the irreducible three-body force is identically zero 
\cite{Sanchis-Alepuz:2015tha}.
The two-body kernel follows, as before, from the quark and quark-gluon vertex 
DSE. 

\section{Results}

\begin{table}[!b]
\centering
\caption{Meson masses and pion decay constant in GeV as calculated in 
rainbow-ladder~\cite{Maris:1999nt} and beyond rainbow-ladder\label{tab:mesons}. 
Results affixed with ${}^\dag$ are fitted values.} 
\setlength{\tabcolsep}{1em}
\begin{tabular}{|c||c|c|c|}
\hline
\hline 
                             &    RL          &       BRL       &     
PDG~\cite{Agashe:2014kda} \\
\hline
\hline                              
       $0^{-+}~(\pi)$        &   $0.14^\dag$  &   $0.14^\dag$   &   
$0.14\phantom{(0)}$    \\
       $0^{++}~(\sigma)$     &   $0.64\phantom{{}^\dag}$       &   
$0.52\phantom{{}^\dag}$        &   $0.48(8)$ \\
       $1^{--}~(\rho)$       &   $0.74\phantom{{}^\dag}$       &   
$0.77\phantom{{}^\dag}$        &   $0.78\phantom{(0)}$    \\
       $1^{++}~(a_1)$        &   $0.97\phantom{{}^\dag}$       &   
$0.96\phantom{{}^\dag}$        &   $1.23(4)$ \\
       $1^{+-}~(b_1)$        &   $0.85\phantom{{}^\dag}$       &   
$1.1\phantom{0^\dag}$         &   $1.23\phantom{(0)}$    \\
\hline       
       $f_\pi$               &   $0.092^\dag$ &   $0.103$  &   $0.092$~~  \\
\hline
\hline 
\end{tabular}
\end{table}
%

%
%
%
%
A consistency check that chiral symmetry is correctly implemented is provided 
by calculating the pion mass in the chiral limit. Finding a massless pion, as 
we indeed do, serves as a verification that the numerics are under 
control.

In~\cite{Williams:2014iea} the internal vertices were adjusted to both resemble the 
tree-level component of the calculated vertex, and to yield a running quark mass in agreement with lattice calculations. Once the 
vertex truncation is defined, the only free parameter left is the 
current-quark mass. The light quark mass is then fixed to $m_{u/d}=3.7$~MeV at the renormalization point $19$~GeV so 
that the physical pion mass is obtained. 

In Table \ref{tab:mesons} we show the ground-state meson masses 
below $1.4$~GeV for two quark flavors for rainbow-ladder (using the Maris-Tandy 
interaction~\cite{Maris:1999nt}) and for the beyond rainbow-ladder truncation 
presented here. The case of the scalar and vector states is remarkable in that 
their masses agree well with the experimental value of the $\rho$- and 
$\sigma$- mesons;  this is, however, an unexpected feature. On the one hand, it 
is well known that unquenching effects (absent in this calculation) 
such as pion-cloud 
corrections~\cite{Oertel:2000jp,Fischer:2007ze,Fischer:2008wy} and 
effects associated with decay 
processes~\cite{Leinweber:2001ac,Allton:2005fb} provide extra attraction for 
vector and/or scalar channels. On the other hand, it is not yet settled whether 
the physical $\sigma$-meson originates (at least in part) form a $q\bar{q}$ state. Moreover, 
even if that was the case, the measured mass would be the result of the 
dynamical coupling of this \emph{bare} state with several other channels 
\cite{Pelaez:2003dy,Doring:2009bi}. In this respect, it is interesting to note 
that several DSE/BSE calculations 
\cite{Fischer:2014xha,Chang:2009zb,Heupel:2014ina}, 
including the present one, generate a 
relatively light scalar $q\bar{q}$ state and hence suggest that the 
picture of a purely dynamical origin of the $\sigma$-resonance is not 
realised; this can likely be traced to the back-coupling of the quark-gluon vertex
to the vertex itself and will be remedied at higher order in loops or in nPI.

The previous remarks can be summarized in that one would expect our 
\emph{quark-core} calculation to lead, in general, to heavier meson masses. 
The fact that this is not the case is an indication that other mechanisms, 
absent in the present calculation, are of relevance even in the 
ground-state meson sector. Additional support for this is found in the 
$a_1/b_1-\rho$ and $a_1-b_1$ splitting; whereas the former is improved with 
respect to the RL value, the latter is 
exceedingly large showing an imbalance in the different 
components of the present quark-gluon vertex.

\begin{table}[!t]
\centering
\caption{Baryon masses in GeV as calculated in rainbow-ladder 
\cite{Eichmann:2009qa,SanchisAlepuz:2011jn,Sanchis-Alepuz:2014mca,Sanchis-Alepuz:2014sca} and beyond rainbow-ladder.\label{tab:baryons}}
\setlength{\tabcolsep}{1em}
\begin{tabular}{|c||c|c|c|}
\hline
\hline
                                 &       RL    &       BRL     &     
PDG~\cite{Agashe:2014kda}       \\
\hline
\hline
    $\nicefrac{1}{2}^+~(N)$      &  $0.94$     &   $1.05$       & 
$0.94\phantom{(0)}$       \\
    $\nicefrac{1}{2}^-~(N^*)$    &  $1.16$     &   $1.08$       & $1.54(1)$    
\\
    $\nicefrac{3}{2}^+~(\Delta)$ &  $1.22$     &   $1.24$       & 
$1.23\phantom{(0)}$       \\ 
\hline
\hline 
\end{tabular}
\end{table}

Let us now turn our attention to baryons. In Table \ref{tab:baryons} we show 
the calculated nucleon and delta-baryon 
masses. We also calculated the mass of the nucleon parity 
partner $\nicefrac{1}{2}^-$, as it is the first signature of the failure of RL 
in the baryon sector. Clearly, the masses for positive-parity 
states are slightly overestimated. This suggests that the simple quark-core 
picture supplemented with pion-cloud effects 
\cite{Oertel:2000jp,Hecht:2002ej,Eichmann:2008ae,Fischer:2008wy,
Sanchis-Alepuz:2014wea}
could be realized for ground-state baryons. The possibility that 
ground-state baryons are, in contrast to mesons, less sensitive to the details 
of the quark-gluon interaction is an interesting one. It explains why the 
simple rainbow-ladder truncation has been thus far so successful in describing 
baryon 
properties~\cite{Eichmann:2009qa,SanchisAlepuz:2011jn,Sanchis-Alepuz:2014mca,
Sanchis-Alepuz:2014sca}.

The situation for excited baryons does not appear to be so simple. In RL 
the negative-parity nucleon comes out extremely light 
\cite{Sanchis-Alepuz:2014mca} unless one inflates the interaction 
strength~\cite{Wilson:2011aa,Segovia:2015hra}, thus spoiling the agreement for 
other observables. The present calculation, using a genuine (albeit truncated) 
solution of a 
 QCD-vertex DSE, does not improve on the situation of the nucleon's 
negative-parity partner. This result by itself already makes a case for the 
systematic introduction of interaction mechanisms, as it is not known \emph{a 
priori} whether 
missing Abelian and/or non-Abelian mechanisms are leading in determining the 
spectrum of (bare) baryon resonances.

Another interesting aspect of using as interaction a solution of the 
quark-gluon vertex DSE is that it naturally features a quark-mass dependence of 
the interaction strength. One class of observables in which such a mass 
dependence is 
manifest are the baryon sigma-terms. Calculated here using the Feynman-Hellmann 
theorem, they measure the dependence of the baryon mass on the quark mass.
As shown in Table~\ref{tab:sigma}, these are exceedingly small in RL, owing to 
the quark-mass independence of the effective interaction. This changes 
dramatically in the present calculation. For the nucleon we now obtain a value 
in reasonable agreement with the upper limit of the consensual range 
\cite{Young:2009zb,Horsley:2011wr,Durr:2011mp,Shanahan:2012wh,Bali:2012qs}. For the delta, although there is no well 
established value to serve as a reference, we find a similarly large result for 
the sigma term. 
 However, this could be related to the absence of decay 
mechanisms that induce non-analytic behaviours in the baryon-mass curve near 
the physical point.

\begin{table}[!t]
\centering
\caption{Sigma terms in MeV as calculated in rainbow-ladder  and beyond 
rainbow-ladder.\label{tab:sigma}}
\setlength{\tabcolsep}{1em}
\begin{tabular}{|c||c|c|c|}
\hline
\hline
                                 &      RL      &      BRL     & 
Other~\cite{Lyubovitskij:2000sf,Cavalcante:2005mb,Erkol:2007sj,Young:2009zb,Horsley:2011wr,Durr:2011mp,Shanahan:2012wh,Bali:2012qs,Flambaum:2005kc} \\
\hline
\hline                                 
    $\nicefrac{1}{2}^+~(N)$      &     $30$     &      $60$    &  $25$--$60$  \\
    $\nicefrac{3}{2}^+~(\Delta)$ &     $24$     &      $63$    &  $32$--$79$    
\\ 
\hline
\hline
\end{tabular}
\end{table}

\section{Summary}

%
%
%
%
In summary, we presented the first calculation of both mesons and baryons in a 
mutually consistent truncation of the quark-gluon vertex beyond rainbow-ladder. 
Being fixed in accordance to the 2PI effective action at three-loops, the freedom
to adjust model parameters is limited and hence the bulk features of the results 
for the meson and baryon masses cannot presently be altered. Extensions must therefore
be made that either take into account higher loop corrections to the effective action,
or as we envisage, to incorporate vertices dynamically into the expansion by considering
3PI actions and beyond~\cite{Sanchis-Alepuz:2015tha}.

Regardless, the means to connect quarks and gluons presented here via
symmetry preserving and calculated interaction vertices is both an important
technical and conceptual achievement that, being in itself independent of
the particular ansatz, can be systematically applied to \emph{e.g.} crossed ladder
and other higher order corrections to the kernel.
Being intrinsically dependent upon the quark flavor and hence the quark mass, it will
form the basis of future investigations of the baryon octet-decuplet~\cite{Sanchis-Alepuz:2014sca} and 
meson nonet for the strange quark~\cite{Qin:2011xq,Fischer:2014xha}, as well as heavy-quark studies of charmonium 
and bottomonium~\cite{Rojas:2014aka,Hilger:2014nma,Fischer:2014cfa}. Furthermore, the impact of corrections beyond rainbow-ladder 
on the internal structure of the hadrons can be tested through the calculation 
of electromagnetic form-factors~\cite{Maris:2002mz,Eichmann:2011vu,Eichmann:2011pv}.

\section*{Acknowledgments}

We thank Gernot Eichmann, Christian S.~Fischer and Walter Heupel for discussions 
and a critical reading of this manuscript. This work has been supported by an 
Erwin Schr\"odinger fellowship J3392-N20 and a Lise-Meitner fellowship 
M1333--N16 from the Austrian Science Fund (FWF), by the Helmholtz International 
Center for FAIR within the LOEWE program of the State of Hesse, and by the DFG 
collaborative research center TR 16.

\end{document}